\documentclass[12pt]{article}
\newcommand{\Purple}[1]{#1}
\newcommand{\Red}[1]{#1}
\newcommand{\Blue}[1]{#1}
\newcommand{\OliveGreen}[1]{#1}
\def\diag{{\rm diag}}
\newcommand{\ft}[2]{{\textstyle\frac{#1}{#2}}}
\def\rmi{{\rm i}}

\csname @addtoreset\endcsname{equation}{section}
\usepackage{epsf,amssymb}
\begin{document}
\begin{titlepage}
\begin{flushright}
KUL-TF-2000/23\\ UG-00-14
hep-th/0010195
\end{flushright}
\vspace{.5cm}
\begin{center}
\baselineskip=16pt
{\LARGE   Symmetries of string, M and F-theories 
}\\
\vfill
{\large Eric Bergshoeff$^1$ and Antoine Van Proeyen $^{2,\dagger }$, 
  } \\
\vfill
{\small
$^1$  Institute for Theoretical Physics, Nijenborgh 4,
9747 AG Groningen, The Netherlands,
\\ \vspace{6pt}
$^2$ Instituut voor
Theoretische Fysica, Katholieke Universiteit Leuven, Celestijnenlaan
200D, B-3001 Leuven, Belgium.
}
\end{center}
\vfill
\begin{center}
{\bf Abstract}
\end{center}
{\small
The $d=10$ type~II string theories, $d=11$ M-theory and $d=12$ F-theory
have the same symmetry group. It can be viewed either as a subgroup of a
conformal group $OSp(1|64)$ or as a contraction of $OSp(1|32)$. The
theories are related by different identifications of their symmetry
operators as generators of $OSp(1|32)$. T- and S-dualities are recognized
as redefinitions of generators. Some $(s,t)$ signatures of spacetime
allow reality conditions on the generators. All those that allow a real
structure are related again by redefinitions within the algebra, due to
the fact that the algebra $OSp(1|32)$ has only one real realization. The
redefinitions include space/space, time/time and space/time dualities. A
further distinction between the theories is made by the identification of
the translation generator. This distinguishes various versions of type~II
string theories, in particular the so-called $*$-theories, characterized
by the fact that the $P_0$ generator is not the (unique) positive-definite
energy operator in the algebra.
\vspace{2mm} \vfill \hrule width 3.cm}
To be published in the proceedings of the G\"{u}rsey Memorial Conference II
\textit{`M-theory and dualities'}, Istanbul, June 2000.\\[5mm]
{\footnotesize
\noindent $^\dagger$ Onderzoeksdirecteur FWO, Belgium }
\end{titlepage}

\section{Introduction}

The group $OSp(1|32)$ was already mentioned in the first papers on $d=11$
supergravity~\cite{CremmerJulia}. This algebra and its extension
$OSp(1|64)$ appeared as anti-de Sitter (adS) and superconformal algebras
in $d=10$ and $d=11$ Minkowski theories~\cite{vanHolten:1982mx} long ago,
and got new attention related to the $M$-theory algebra~\cite{Malgebra}.
The adS or conformal algebras got new attention in a recent paper on the
superconformal aspects of $d=11$ theories \cite{West:2000ga} and in
two-time theories \cite{BDM2t,BDMliftM}. In these two cases, the
$OSp(1|64)$ conformal group appeared. In the physical theories that we
consider, we need the subgroup of $OSp(1|64)$ that is a contraction of
$OSp(1|32)$ in a way that will be clarified below.

Our initial motivation to study the role of the $OSp(1|32)$ algebra was
related to Euclidean theories. When one considers the $D$-instanton
\cite{Dinstanton}, one often considers the bosonic theory, ignoring its
possible embedding in the supersymmetric theory. In particular, one makes
use of the IIB theory in Euclidean space, while the latter can not be
formulated as a supersymmetric theory with real fields, as we will show
below. Remark that the connection between these Euclidean theories and
the Minkowski string theories involve a duality between theories of
different spacetime signature \cite{dualst}.

A second question that was posed when we started this research, was
related to the observation that in many super-Euclidean theories one makes
use of complexification of the fields and in other cases one does not
\cite{Euclidean}. We would like to know when it is necessary to do so, and
when it can be avoided.

Apart from the possibility of no time directions, one is also interested
in theories with more time directions
\cite{BarsS,Hull,BDM2t,BDMliftM,StarTheories,Bars2tFT}. Therefore, it
looked natural to extend our investigation to an arbitrary spacetime
signature.

This leads to a web of dualities between theories in $d=10$, 11 and 12 of
different spacetime signature, similar to what has been found in
\cite{Hull}. We obtain these dualities from an algebraic approach, which
puts the contraction of $OSp(1|32)$ as a unifying principle. The different
theories are then just many faces of the same underlying symmetry group.
This seminar summarizes the results obtained in \cite{faces}.
\par
In section~\ref{ss:PoinAdSConf}, we clarify the relation between the
super-Poincar\'{e} algebra that we consider here and the full $OSp(1|32)$ as
super-adS algebra or $OSp(1|64)$ as superconformal algebra. Then we go to
the main 3 steps of our results. In section~\ref{ss:complexalgebras} we
consider the complex algebra and its realizations in the different
dimensions, in section~\ref{ss:realalgebras} we discuss the real algebra
and its realizations in different spacetime signatures, and in
section~\ref{ss:TranslationEnergy} we identify the translation operator,
distinguishing between different Lagrangian theories for the same
spacetime signature. Throughout the work we indicate the dualities
connecting all the theories. Finally, a short summary is given in
section~\ref{ss:conclusions}. Some extra figures are given in
\cite{esfFaces}.

\section{Poincar\'{e} algebras as contractions of anti-de Sitter and as
subalgebras of conformal algebras}\label{ss:PoinAdSConf}

The Poincar\'{e} algebra contains translations and the Lorentz algebra.
\begin{eqnarray}
\left[ \Red{P_\mu},\Red{P_\nu}\right]&=& 0\,,\qquad
\left[\Red{P_\mu} , \Blue{M_{\nu\rho}}\right]=\eta_{\mu[\nu}\Red{P_{\rho]}}\,,\nonumber\\
\left[\Blue{M_{\mu\nu}},\Blue{M_{\rho\sigma}}\right]
&=&\eta_{\mu[\rho}\Blue{M_{\sigma]\nu}}
-\eta_{\nu[\rho}\Blue{M_{\sigma]\mu}}\,. \label{Poinalgebra}
\end{eqnarray}
It is a semi-direct product of $SO(d-1,1)$ with translations.

In the adS algebra, the translations unify with the rotations to
$SO(d-1,2)$. The translations do not commute anymore,
\begin{equation}
  \left[ \Red{P_\mu},\Red{P_\nu}\right]= \frac{1}{2
  \Purple{R}^2}\Blue{M_{\mu\nu}}\,,
 \label{adSPP}
\end{equation}
where a parameter $\Purple{R}$ appears that is the radius of the adS
space. The Poincar\'{e} algebra is the contraction of the algebra
(\ref{adSPP}) obtained by $\Purple{R}\rightarrow \infty$. If the
right-hand side of (\ref{adSPP}) would be $-\frac{1}{2
\Purple{R}^2}\Blue{M_{\mu\nu}}$, we would have the \OliveGreen{de Sitter
algebra}, rather than the \OliveGreen{anti-de Sitter ($adS$) algebra}.
The structure of the algebra is clarified by defining
\begin{equation}
 M_{d\mu}\equiv -M_{\mu d}\equiv {\Purple{R}}\,\Red{P_\mu}\,,
\label{defMdmu}
\end{equation}
to obtain the algebra
\begin{equation}
\left[M_{\hat\mu\hat\nu},M_{\hat\rho\hat\sigma}\right]
=\eta_{\hat\mu[\hat\rho}M_{\hat\sigma]\hat\nu}
-\eta_{\hat\nu[\hat\rho}M_{\hat\sigma]\hat\mu}\,, \label{adSSO}
\end{equation}
where $\hat\mu=0,\ldots ,d$, and $\eta_{\hat\mu\hat \nu}=\diag (-+\ldots
+-)$ (for de Sitter, rather than adS, the latter $-$ would be another
$+$).

The conformal algebra is $SO(d,2)$ having the Poincar\'{e} algebra as a
subalgebra (the first two lines of these equations):
\begin{eqnarray}
\left[ \Red{P_\mu},\Red{P_\nu}\right]&=& 0\,,\qquad
\left[\Red{P_\mu} , \Blue{M_{\nu\rho}}\right]=\eta_{\mu[\nu}\Red{P_{\rho]}}\,,\nonumber\\
\left[\Blue{M_{\mu\nu}},\Blue{M_{\rho\sigma}}\right]
&=&\eta_{\mu[\rho}\Blue{M_{\sigma]\nu}}
-\eta_{\nu[\rho}\Blue{M_{\sigma]\mu}}\,,\nonumber\\
\left[ K_{\mu} , \Blue{M_{\nu\rho}}\right] & =& \eta_{\mu[\nu}K_{\rho]}
\,, \qquad
\left[ K_\mu ,K_\nu \right] =0\,,\nonumber\\
\left[\Red{P_{\mu}} , K_{\nu}\right]  &=& 2 (\eta_{\mu\nu} \Purple{D} + 2
\Blue{M_{\mu\nu})}\,,
\nonumber\\
\left[ \Purple{D},\Red{P_\mu}\right]  &=&\Red{P_\mu}\,,\qquad \left[
\Purple{D},\Blue{M_{\mu \nu }}\right]  =0\,,\qquad
 \left[ \Purple{D},K_\mu\right]  =-K_\mu\,. \label{algConform}
\end{eqnarray}
The commutation relations with the dilations in the last line define a
weight for all the generators, giving a weight~1 to $P$, weight~0 to
$M_{\mu \nu }$. All the commutation relations are consistent with the
weight assignments. In this way the algebra is visually represented by the
diagram
\begin{eqnarray}
 1 & : & P_\mu  \nonumber\\
 0 & : & D,\ M_{\mu \nu }\nonumber\\
 -1 & : & K_\mu \,.
 \label{weightsConf}
\end{eqnarray}
This is related to the 3-graded structure, which for the superalgebras
will be the 5-graded structure, that Murat G\"{u}naydin was mentioning at this
conference.

\begin{figure}
\begin{center}
\leavevmode \epsfxsize=8cm
 \epsfbox{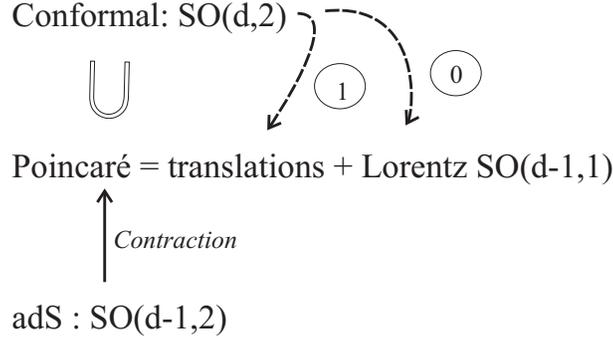} 
\caption{\it Schematic structure of bosonic algebras\label{fig:strAlg}.
The numbers in circles denote the weights discussed in the text.}
\end{center}
\end{figure}

Schematically, the structure of the algebras that we have encountered
contains a contraction of an adS algebra, and an inclusion of the
Poincar\'{e} algebra in a conformal algebra, see figure~\ref{fig:strAlg}.
Note that the Poincar\'{e} algebra is built from the weight~1 generators and
the Lorentz generators in the weight~0 part of the conformal algebra. The
weight~$-1$ generators are realized non-linearly in the physical theories.

\begin{figure}
\begin{center}
\leavevmode \epsfxsize=10cm
 \epsfbox{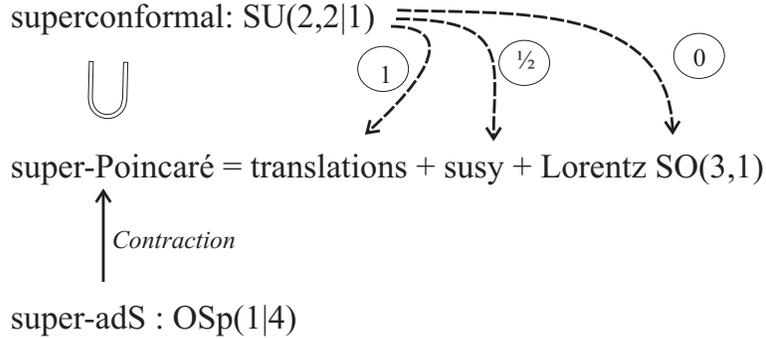}
\caption{\it Schematic structure of superalgebras for $N=1$,
  $d=4$.\label{fig:supalgebrad4}}
\end{center}
\end{figure}

For the superalgebras (we take 4 dimensions and $N=1$), the schematic
picture looks similar (see figure~\ref{fig:supalgebrad4}).
 The adS algebra that is at
the basis, is $OSp(1|4)$. Rescaling the $P$ generators to $P/x$,
and the supersymmetries $Q$ to $Q/\sqrt{x}$, the superalgebra is%
\footnote{We omit spinor indices, as well as the charge conjugation
matrix ${\cal C}^{-1}$ that should be multiplied with the gamma matrices
on the right-hand side of all anticommutation relations. This charge
conjugation can be seen as the metric that lowers the (unwritten) spinor
indices.}
\begin{eqnarray}
 \left\{ \Red{Q},\Red{Q}\right\} & = & \gamma ^\mu \Red{P_\mu }+ 2\Purple{x}
 \gamma ^{\mu \nu }\Blue{M_{\mu \nu }} \,,\qquad
 \left[\Red{P_\mu },\Red{Q}\right]    =  \Purple{x}\gamma _\mu \Red{Q} \,,\nonumber\\
 \left[ \Blue{M_{\mu \nu }},\Red{Q}\right] &=& -\ft14 \gamma _{\mu \nu }\Red{Q}\,,\qquad
 \left[ \Red{P_\mu },\Red{P_\nu }\right] = 8\Purple{x^2}\Blue{M_{\mu \nu }}\,,\nonumber\\
\left[\Red{P_\mu} ,
\Blue{M_{\nu\rho}}\right]&=&\eta_{\mu[\nu}\Red{P_{\rho]}}\,,\qquad
\left[\Blue{M_{\mu\nu}},\Blue{M_{\rho\sigma}}\right]
=\eta_{\mu[\rho}\Blue{M_{\sigma]\nu}}
-\eta_{\nu[\rho}\Blue{M_{\sigma]\mu}}\,.
\end{eqnarray}
The super-Poincar\'{e} theory is the contraction $x\rightarrow 0$ of this
algebra.

When we consider the superconformal algebras, the dilatational weights of
the generators form a table
\begin{eqnarray}
 1 & : & P_\mu  \nonumber\\
 \ft12 & : & Q \nonumber\\
 0 & : & D,\ M_{\mu \nu },\ R\nonumber\\
 -\ft12 & : & S \nonumber\\
 -1 & : & K_\mu \,,
\label{weightsSCAlg}
\end{eqnarray}
where $R$ is for this case $U(1)$, and in general can be a larger
`$R$-symmetry' algebra. The super-Poincar\'{e} subalgebra is obtained from the
positive weight operators, and the Lorentz part of the weight~0
operators. The others can appear non-linearly realized.

\begin{figure}
\begin{center}
\leavevmode \epsfxsize=12cm
 \epsfbox{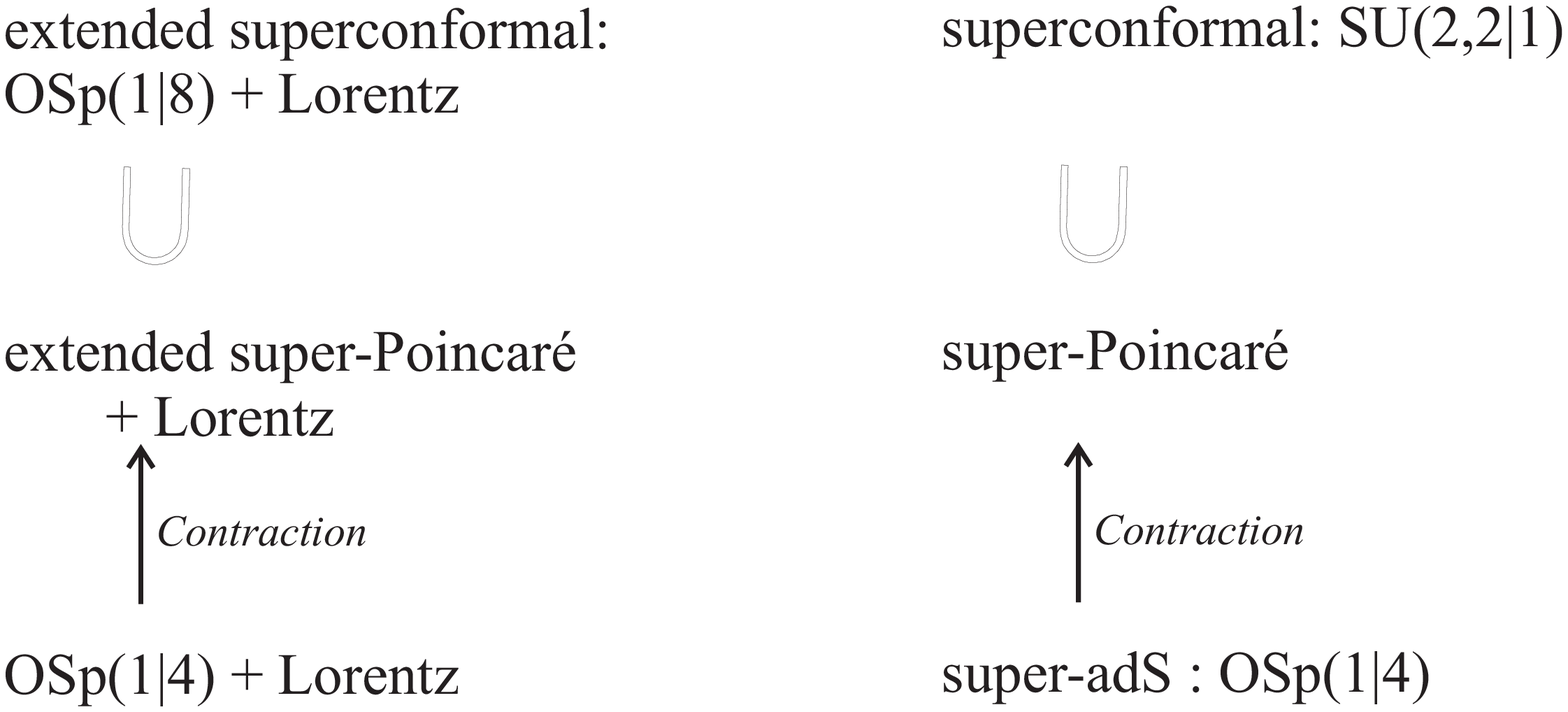}
\end{center}\vspace{-3.5cm}
\phantom{.}\hfill $\left\{ \Red{Q},\Red{Q}\right\}  =  \gamma ^\mu \Red{P_\mu }$\\[2cm]
 $\begin{array}{l}
\left\{ \Red{Q},\Red{Q}\right\}  =  \gamma ^\mu \Red{P_\mu }+ 2
  \gamma ^{\mu \nu }\Red{Z_{\mu \nu }}\qquad\qquad \qquad
  \left\{ \Red{Q},\Red{Q}\right\}  =  \gamma ^\mu \Red{P_\mu }+ 2 \gamma ^{\mu \nu }\Blue{M_{\mu \nu
   }}\\
   M_{\mu \nu } \end{array}$
\centering\caption{\it (Extended) algebras for $N=1$,
  $d=4$.\label{fig:extalgebrad4}}
\end{figure}

There is an enlargement of the picture, see figure~\ref{fig:extalgebrad4}.
Indeed, the 2-index operator that appears on the right-hand side of the
supersymmetry anticommutators, should not necessary be identified with
the Lorentz generators. We can give it the name $Z_{\mu \nu }$,
suggesting its identification as a `central' charge:
\begin{equation}
   \left\{ \Red{Q},\Red{Q}\right\}  =  \gamma ^\mu \Red{P_\mu }+ 2 \gamma ^{\mu \nu }\Blue{M_{\mu \nu
   }}\,.
 \label{QQPZ}
\end{equation}
The Lorentz generators $M_{\mu \nu }$ are then an extra part of the
algebra. Thus, there is a semi-direct sum of the Lorentz algebra and the
algebra $OSp(1|4)$. Now we apply a different type of contraction. First,
we rescale also $Z_{\mu \nu }$ in the same way as $P_\mu $ (remember that
$M_{\mu \nu }$ was not rescaled), arriving at
\begin{eqnarray}
 \left\{ \Red{Q},\Red{Q}\right\} & = & \gamma ^\mu \Red{P_\mu }+ 2
  \gamma ^{\mu \nu }\Red{Z_{\mu \nu }} \,,\qquad
 \left[\Red{P_\mu },\Red{Q}\right]    =  \Purple{x}\gamma _\mu \Red{Q} \,,\nonumber\\
 \left[ \Red{Z_{\mu \nu }},\Red{Q}\right] &=& -\ft14\Purple{x} \gamma _{\mu \nu }\Red{Q} \,,\qquad
 \left[ \Red{P_\mu },\Red{P_\nu }\right] = 8\Purple{x^2}\Red{Z_{\mu \nu }} \,,\nonumber\\
\left[\Red{P_\mu} ,
\Red{Z_{\nu\rho}}\right]&=&\Purple{x}\eta_{\mu[\nu}\Red{P_{\rho]}}
\,,\qquad \left[\Red{Z_{\mu\nu}},\Red{Z_{\rho\sigma}}\right]
=\Purple{x}\eta_{\mu[\rho}\Red{Z_{\sigma]\nu}}
-\Purple{x}\eta_{\nu[\rho}\Red{Z_{\sigma]\mu}}\,.
\end{eqnarray}
The rescaling in this case keeps the $Z$ in the anticommutator of the
supersymmetries, as in (\ref{QQPZ}). The extended super-Poincar\'{e} algebra
contains thus also the central charge, and the Lorentz generators as extra
part. This is a subalgebra of the extended superconformal algebra
$OSp(1|8)$, which was already mentioned in \cite{vanHolten:1982mx} as a
second possible $N=1$, $d=4$ superconformal algebra.

\begin{figure}
\begin{center}
\leavevmode \epsfxsize=12cm
 \epsfbox{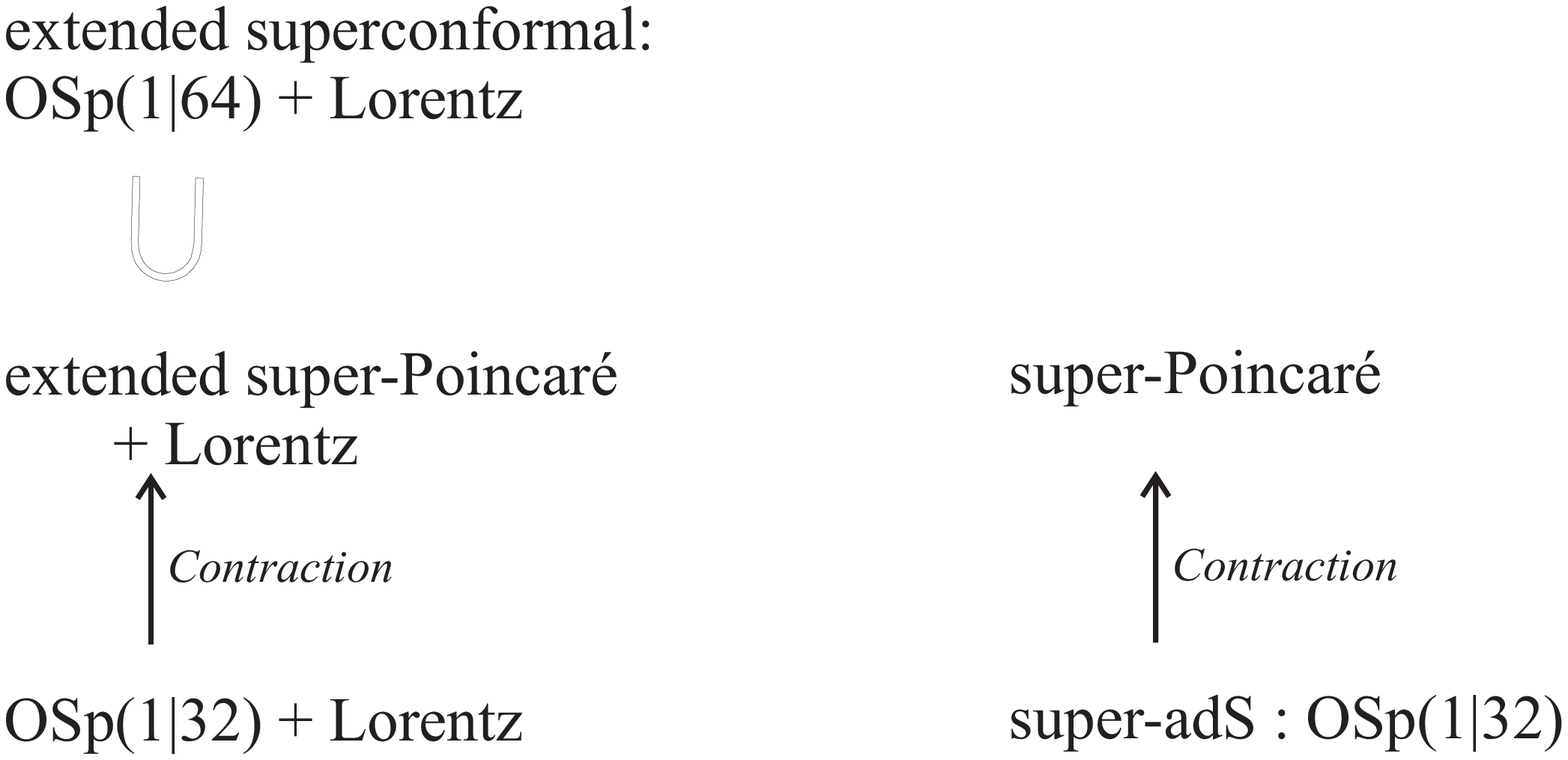} 
\caption{\it (Extended) algebras for $d=11$.\label{fig:extalgebrad11}}
\end{center}
\end{figure}

A similar picture can be made e.g. in $d=11$, see
figure~\ref{fig:extalgebrad11}.
The adS algebra that forms the basis is then $OSp(1|32)$. The
anticommutator of the supersymmetries contains, apart from the
translations, a 2-index and a 5-index generator:
\begin{equation}
   \left\{ Q_\alpha,\,Q_\beta\right\}=
  \Gamma^\mu _ {\alpha\beta} P_\mu   +
 2\Gamma^{\mu\nu}_{\,\alpha\beta} Z_{\mu\nu}
+\frac{1}{5!}\Gamma^{\mu\nu\rho\sigma\tau}_{\,\alpha\beta}
Z^5_{\mu\nu\rho\sigma\tau}\,.
 \label{QQd11}
\end{equation}
One may use uniform rescalings for all the bosonic generators (dividing
them by $x$) and the fermionic $Q$ (dividing them by $\sqrt{x}$). Then
the anticommutator of the supersymmetries is not affected, but all the
other commutators ([boson,boson] and [fermion,boson]) get a factor $x$ on
the right-hand side, and thus vanish in the corresponding Poincar\'{e}
contracted theory. One may also identify the 2-index operator as the
Lorentz generator, as it has the appropriate commutation relations with
the supersymmetries and with the bosonic generators. If this one is not
rescaled by a factor $x$, the contracted theory does not contain $M_{\mu
\nu }$ in the right-hand side of the anticommutator of supersymmetries,
and the appropriate commutators with the Lorentz generators do survive
the limit. In this case, there is, however, only one superconformal
algebra \cite{vanHolten:1982mx}. The positive weight operators of that
algebra do contain a $Z_{\mu \nu }$, and thus it is the extended
super-Poincar\'{e} algebra that is a subgroup of this superconformal algebra.

Below, we will be considering this extended super-Poincar\'{e} algebra,
having as generators those of $OSp(1|32)$. It is clear from the above,
that we could also consider it as a subalgebra of $OSp(1|64)$. The latter
contains also the  generators of negative weight that are important for
the nonlinear realizations as in \cite{West:2000ga}. This algebra would
also allow to make connections to higher dimensions
\cite{BarsS,BDMliftM}. But the extra generators are not relevant for the
issues that we treat here. Only the anticommutator of 32 supersymmetries,
the only non-trivial one in the super-Poincar\'{e} theory, will be mentioned.

\section{Complex symmetry algebras}\label{ss:complexalgebras}

$OSp(1|32)$ is the algebra of 32 fermionic charges with all possible
bosonic generators in their anticommutator. The contraction, in the sense
indicated in section~\ref{ss:PoinAdSConf}, underlies the F-theory of 12
dimensions, the M-theory of 11 dimensions, and the IIA and IIB string
theories in 10 dimensions. They are obtained by identifying appropriate
subgroups of $Sp(32)$ as the Lorentz rotations. Note that in the case of
the extended algebras of section~\ref{ss:PoinAdSConf}, this $Sp(32)$ is
the automorphism algebra of the supersymmetries, in the semi-direct
product with $OSp(1|32)$. In any case, the supersymmetries should be in a
spinor representation of the Lorentz group. Dimensional reduction and
T-dualities are then obtained as mappings between generators of
$OSp(1|32)$.

To recognize $OSp(1|32)$ as a symmetry algebra in $d$ dimensions, one has
to embed $SO(d)$ in $Sp(32)$, in such a way that the spinor
representation of $SO(d)$ fits in the 32. This makes already clear that
$d=12$ is the highest possible dimension. To make that identification, we
have to select chiral spinors $\Red{\hat{Q}}$ of 12 dimensions. These are
defined using the chiral projection $\hat{{\cal P}}^+$:
\begin{equation}
\hat{{\cal P}}^+  \Red{\hat{Q}}= \Red{\hat{Q}}\,,\qquad \hat{{\cal
P}}^+=\ft12(1+\hat{\Gamma} _*)\,,\qquad \hat{\Gamma} _*=\Gamma _1\Gamma
_2\ldots \Gamma_{12}\,.
 \label{chirald12}
\end{equation}
Remark that we use the notation $\Gamma _*$ (the hat specifies the
12-dimensional context) in any even dimension to denote the product of all
the gamma matrices, similar to $\gamma _5$ in 4 dimensions.
The anticommutator of the supersymmetries looks like
\begin{equation}
   \left\{\Red{\hat{Q}},\Red{\hat{Q}} \right\} = \ft12 \hat{{\cal P}}^+ \hat{\Gamma
}^{\hat{M}\hat{N}}\Blue{\hat{Z}_{\hat{M}\hat{N}}}  +\ft1{6!} \hat{{\cal
P}}^+ \hat{\Gamma }^{\hat{M}_1\cdots
\hat{M}_6}\Blue{\hat{Z}^+_{\hat{M}_1\cdots \hat{M}_6}}\,.
\label{algd12}
\end{equation}

In 11 dimensions, the bosonic generators split as $528=11+55+462$,
following the anticommutator (\ref{QQd11}). In 10 dimensions one can
again define chiral spinors, which are 16-dimensional, and consider either
2 generators of opposite chirality (IIA) or of the same chirality (IIB).
In the first case, the anticommutators are
\begin{eqnarray}
\left\{Q^{\pm },Q^{\pm } \right\} & = & {\cal P}^{\pm }\Gamma ^M Z^{\pm
}_M+\ft 1{5!}{\cal P}^{\pm }\Gamma ^{M_1\cdots M_5}Z^\pm _{M_1\cdots
M_5}\, ,
\nonumber\\
\left\{Q^{\pm },Q^{\mp } \right\} & = & \pm {\cal P}^{\pm } Z +
\ft12{\cal P}^{\pm }\Gamma ^{MN}Z_{MN} \pm \ft 1{4!}{\cal P}^{\pm }\Gamma
^{M_1\cdots M_4}Z _{M_1\cdots M_4}\,. \label{calgIIA}
\end{eqnarray}
The 528 generators are thus split as $2\times (10+126)$ in the
anticommutators between generators of the same chirality and $1+45+210$
in the anticommutator between generators of opposite chirality.

For the IIB case, we have a doublet of fermionic generators $Q^i$,
$(i=1,2)$, of the same chirality, and the anticommutators are
\begin{eqnarray}
  \left\{Q^i,Q^j \right\}&=&{\cal P}^+\Gamma ^MY_M^{ij}
     +\ft1{3!}{\cal P}^+\Gamma ^{MNP}\varepsilon ^{ij}Y_{MNP}
  +{\cal P}^+\ft1{5!}\Gamma ^{M_1\cdots M_5 }Y_{M_1\cdots M_5 }^{+\,ij}\,,
\nonumber\\
Y_M^{ij} & = &  \delta ^{ij}Y_M^{(0)}
   +\tau _1^{ij}   Y_M^{(1)}+\tau _3^{ij}   Y_M^{(3)}\,,
\label{calgIIB}
\end{eqnarray}
where in the second line we have split the symmetric matrix $Y^{ij}$ in
three components, as we can also do for the 5-index generators. The
decomposition is here $528=(3\times 10)+120+(3\times 126)$.

It is clear that all these algebras are related. The dimensional
reductions relate the generators as follows. The chiral generator
$\hat{Q}$ of 12 dimensions, splits in 10 dimensions in a chiral and an
antichiral generator, as follows from the relation $\hat{\Gamma
}_*=\Gamma _*\otimes \sigma _3$ for a convenient realization of gamma
matrices, where $\Gamma _*$ is the product of 10 gamma matrices of
dimension $32\times 32$ in 10 dimensions (for the realization that we use
in any dimension see \cite{tools}). The two chiral generators are the
$Q^\pm $ in (\ref{calgIIA}), and adding them gives the 32-component
generator $Q=Q^++Q^-$ used for $d=11$.  The T-dual theories are
identified by taking
\begin{equation}
  Q^+= Q^1\,,\qquad Q^-=\Gamma^sQ^2\,,
 \label{QTdual}
\end{equation}
where $\Gamma ^s$ is a gamma matrix in an arbitrary (spacelike or
timelike) direction. On the other hand, S-duality is the mapping
\begin{equation}
  Q^i  \stackrel{S}{\longrightarrow}  \left( e^{\rmi\ft14 \pi \tau _2}\right)
{}^i{}_jQ^j\,.
 \label{QSdual}
\end{equation}
Thus all the dimensional reductions and dualities are written as mappings
between the generators of $OSp(1|32)$. We  mentioned here only the
fermionic generators explicitly, as the rules for the bosonic generators
follow from identifying the anticommutator relations before and after the
map.

\section{Real symmetry algebras}\label{ss:realalgebras}

The important fact for the real forms is the uniqueness of the real form
of the superalgebra $OSp(1|32)$. Therefore the equivalences of all the
symmetry algebras of section~\ref{ss:complexalgebras} are valid also for
the real form, when it exists. The real form  exists only for specific
spacetime signatures. The dimensional reduction and T-duality acts now
between theories of specific signatures. We have to distinguish then
space/space, time/time and space/time dualities.

In general, a complex algebra has different real forms. Even for an
algebra as small as $SU(2)$,
\begin{equation}
   [\Red{T_1},\Red{T_2}]=\Red{T_3}\,,\qquad
   [\Red{T_2},\Red{T_3}]=\Red{T_1}\,,\qquad
  [\Red{T_3},\Red{T_1}]=\Red{T_2}\,,
 \label{SU2c}
\end{equation}
there are already two inequivalent `real forms'. Either one considers the
set $\Blue{a^1} \Red{T_1} +\,\Blue{a^2} \Red{T_2}+ \Blue{a^3}\Red{T_3}$
with $\Blue{a^i}\in \mathbb{R}$, which is the real form $SU(2)$, or one
can consider $\rmi \Blue{b^1} \Red{T_1} +\rmi \Blue{b^2} \Red{T_2}+
\Blue{b^3}\Red{T_3}=\Blue{b^i}\Red{S_i}$, with $\Blue{b^i}\in \mathbb{R}$,
for $S_{1}=\rmi T_1$, $S_2=\rmi T_2$ and $S_3=T_3$. Another way of saying
this is that either the $T$ or the $S$ generators can be considered as
real. In both cases no $\rmi$ appear in the commutation relations, see
\begin{equation}
   [\Red{S_1},\Red{S_2}]=-\Red{S_3}\,,\qquad
   [\Red{S_2},\Red{S_3}]=\Red{S_1}\,,\qquad
  [\Red{S_3},\Red{S_1}]=\Red{S_2}\,,
 \label{SU11}
\end{equation}
which differs with one minus sign from (\ref{SU2c}). The minus sign can
not be eliminated by real redefinitions. The fact that they are the same
complex algebra means that they are the same by complex redefinitions.

Considering the table of real forms of the basic Lie superalgebras `of
classical type' \cite{realLieSA} (see e.g.  table~5 of \cite{tools} for a
convenient presentation), we see that nearly all superalgebras have
different real forms, even the exceptional superalgebras. But the
algebras $OSp(1|n)$ have only one real form, with $Sp(n,\mathbb{R})$ as
bosonic subalgebra.

To realize this real algebra in $d$ dimensions, we have to consider when
we can impose consistent reality conditions on the fermionic generators.
This is sufficient to be able to classify all the realizations of the
unique real superalgebra $OSp(1|32)$. This has been investigated in
\cite{Gammamatrices}, and table~2 of \cite{tools} gives the summary of
the results that we need. We need 32 real supercharges. The table shows
that $d=12$ with $(space,time)$ signature $(10,2)$ is the highest
possible dimension. In general the results are invariant under
$(s,t)\simeq (s-4,t+4)$, thus $(6,6)$ is also possible. The interchange
of $s$ and $t$ is irrelevant, and corresponds merely to a change of
notations of mostly $+$ to mostly $-$ metrics. Therefore we do not
mention the $(2,10)$ signature. To make the projections to real spinors
one uses three types of projections, Weyl, Majorana or symplectic
Majorana. This leads to the possibilities for 32-component real spinors
listed in table~\ref{tbl:MWSspinors}.
\begin{table}[ht]
\tabcolsep 2pt
\begin{center}
\begin{tabular}{|c|cccccccccccc|}
\hline
12  & &&(10,2) &  &  &  & &&&& (6,6) &  \\
64  & &&MW     &  &  &  & &&&&MW &  \\[3mm]
11  & &(10,1) & & (9,2) &  & \phantom{(9,2)} &&\phantom{(9,2)}&& (6,5)&=&(5,6)   \\
32  & &M      & & M     &  &  &&&&  M &&M  \\[3mm]
10  & (10,0) && (9,1)  && (8,2)&& (7,3) && (6,4) && (5,5)& \\
32  & SM     && MW     && M    && SMW   &&  SM   && MW &\\
    & A      && A/B    && A    &&  B    &&  A    && A/B&\\
\hline
\end{tabular}\tabcolsep 6pt
  \caption{\sl The possible spacetime signatures for 32 real spinor generators.
The first column indicates the number of complex generators that are
present before any projection. The last row indicates, for each
signature, whether in $d=10$ a real form for type~IIA (A), type~IIB (B)
or both (A/B) exists.
 }\label{tbl:MWSspinors}
\end{center}
\end{table}

One can then consider the dimensional reductions and T-dualities
discussed in section~\ref{ss:complexalgebras}, but now we have to be
careful with the signatures. When performing the T-duality as in
(\ref{QTdual}), one has to distinguish whether $\Gamma ^s$ is a timelike
or a spacelike gamma matrix. This $s$-direction can even be timelike for
the IIA algebra and spacelike for the IIB algebra or vice--versa. These
are the time/space or space/time T-dualities, changing the signature.
This leads to the both-sided arrows in figure in table~2 of \cite{faces}.

\section{Translations and the energy operator}\label{ss:TranslationEnergy}

In the third step we identify one of the vector generators as
`translations'. This identification is essential for a spacetime
interpretation of the theory. The different possibilities for this
identification distinguish e.g. IIA from IIA$^*$ theories. We will then
remark that T-duality gives a mapping between different types of
generators. It can mix e.g. translations with `central charges'. Finally,
we will see that there is a unique positive energy operator in the
algebra. However, that generator is not always the timelike component of
the translations. For instance, in IIA theories, the positive operator is
$P_0$, but in IIA$^*$ theories it is another one, and thus $P_0$ is not
positive in that case.

So far, all bosonic generators were treated on equal footing. To make the
connection between algebras and a spacetime theory, we want to know which
generator performs `translations' in spacetime. Seen in another way,
spacetime is the manifold defined from a base point by the action of this
`translation' generator. This is thus similar to the coset space idea. To
generate a spacetime of the appropriate dimension, the translation
operator should be a vector operator in the theory. This is nearly the
only requirement, apart from a non-degeneracy condition. Indeed, in order
that the supersymmetries perform their usual role, they should square to
the translations. Thus the matrix that appears in the anticommutator
between all the supersymmetries, defining how they square to
translations, should be non-degenerate.

For $d=12$, with the algebra (\ref{algd12}), there is no vector operator.
Thus there is no candidate for translations, implying that F-theory has
no straightforward spacetime interpretation. On the other hand, for
$d=11$, with the algebra (\ref{QQd11}), there is one vector operator, and
this one should thus be called the translation generator.

In 10 dimensions it becomes more interesting. Consider first the IIA
algebra (\ref{calgIIA}). There are 2 vector operators $Z_M^+$ and
$Z_M^-$. Both separately are not convenient, because then one half of the
supersymmetries would not square to translations. But we can take linear
combinations. For the signature $(9,1)$ there are, up to redefinitions,
two choices consistent with the reality conditions
\begin{eqnarray}
(9,1)\ :\ {\rm IIA}\phantom{*} & : & \Purple{P_M}\equiv \Red{Z_M^+} + \Red{Z_M^-}\,, \nonumber\\
{\rm IIA}^* & : &\Purple{P_M}\equiv \Red{Z_M^+} - \Red{Z_M^-}\,.
\label{defIIA*}
\end{eqnarray}
We label these choices as IIA and IIA$^*$ in accordance with \cite{Hull}.
For signatures $(10,2)$ or $(8,2)$ there are the possibilities
\begin{eqnarray}
(10,0)\mbox{ or }(8,2)\ :\ \textrm{IIA}& : &
\Purple{P_M}\equiv \rmi(\Red{Z_M^+} + \Red{Z_M^-})\,,\nonumber\\
  \textrm{IIA}'& :& \Purple{P_M}\equiv \Red{Z_M^+} -   \Red{Z_M^-}\,.
\label{defPIIAp}
\end{eqnarray}
However, now these two choices can be related by a redefinition $Q^\pm
\rightarrow e^{\pm \rmi\pi /2} Q^\pm$. Such a redefinition is, similar to
(\ref{QSdual}) and therefore we also recognize it as an S-duality.

The operators that are not translations, remain as `central charges' in
the theory. Therefore we see that the generator that is a translation in
one theory, appears as a central charge in the other theory, as we
announced in the beginning of this section.

For the IIB algebra (\ref{calgIIB}), there are three candidates for
translations. For signature (9,1) they are all consistent with the
reality condition. We thus distinguish
\begin{eqnarray}
(9,1)\ :\hskip .2truecm  && {\rm IIB}\phantom{*} \ : \
\Purple{P_M}=\Red{Y^{(0)}_M}\, , \nonumber\\
&&
{\rm IIB}^* \  : \ \Purple{P_M}=\Red{Y^{(3)}_M}\,, \nonumber\\
&&{\rm IIB}' \  : \ \Purple{P_M}=\Red{Y^{(1)}_M}\,.
 \label{PIIB}
\end{eqnarray}
Considering the possible redefinitions, we come back to (\ref{QSdual}).
This S-duality leaves the IIA translation generator invariant, and
relates the translation of IIB$^*$ with that of ${\rm IIB}'$. On the other
hand, for the signature (7,3) there are three S-dual versions.

Finally, we can identify one bosonic operator in $OSp(1|32)$ that is
positive. We can identify this one from the anticommutation relation
\begin{equation}
  \left\{Q^i, Q^{j\,*}\right\}  \geq
   0\,,
 \label{PositiveQQ}
\end{equation}
using the Majorana condition. Denoting
\begin{equation}
  \left\{Q^i, Q^{j}\right\}=  \mathcal{Z}^{ij} {\cal C}^{-1}
 \label{QQZC}
\end{equation}
to represent all the anticommutation relations, where $\mathcal{Z}^{ij}$
is a matrix in spinor space as well as in the $i,j$ indices, this implies
\begin{equation}
  \mathcal{Z}^{ij}\Gamma ^t \cdots \Gamma ^1 \geq 0\,,
 \label{Zpos}
\end{equation}
exhibiting all the timelike $\Gamma $-matrices. Therefore, the trace of
that operator has positive eigenvalues. When we split $Z$ as usual in
different irreducible representations for the spacetime Lorentz group,
then, in order to absorb the gamma matrices, the relevant part of $Z$ has
as many spacetime indices as there are time directions. All its
directions should be timelike. Thus for Minkowski spaces it is the
timelike component of a vector operator, while for Euclidean theories
this positive operator is a scalar `central charge'. For Minkowski
theories we can thus wonder whether the positive energy operator is the
timelike component of the operator that we selected as `translations'. If
this is the case then the usual Hamiltonian will be positive. When the
positive energy is the timelike component of another vector operator,
then the Hamiltonian built from $P_0$ is not positive definite. This is
what happens in the IIA$^*$, IIB$^*$ and IIB$^\prime$ theories. In these
theories, the kinetic energies of some of the $p$-form gauge fields are
negative definite. As an example consider the vector operators in the
IIB-like theories, as in (\ref{calgIIB}). With our convention that
$\alpha =1$, the trace in (\ref{Zpos}) selects the $M=0$ component of
$\Red{Y_M^{(0)}}$ as the positive definite energy. Thus it is indeed the
IIB theory where the energy is the timelike part of translations, and not
for the other versions. The algebraic approach thus gives an
understanding of the positivity in type IIA and IIB versus lack of
positivity in the other theories.

\section{Conclusions}\label{ss:conclusions}

The algebras of F-theory, M-theory, type IIA and IIB, ... are different
faces of the same superalgebra $OSp(1|32)$. The uniqueness of the real
form of that algebra implies that all these manifestations can be related
by mappings between the  generators of the algebra. That holds especially
for the dimensional reductions, T- and S-dualities that relate these
theories. Different spacetime signatures are easily incorporated.
However, for certain spacetime signatures, some theories may exist only
in complex form. That answers the questions about why we need sometimes a
complexification procedure to obtain an Euclidean theory. In particular,
the IIB theory has no real form in (10,0). Therefore, in order to discuss
the D-instanton in IIB, we have to give up the concept of a theory with
real fields and action.

We have understood the *-theories as being distinguished from the usual
IIA and IIB by a different identification of the translation generator.
They are related to the ordinary theories by a `duality' interchanging
translations with central charges. Thus in these dualities the concept of
spacetime is very intriguing. It should be interchanged with a sort of
harmonic space where coordinates are associated also with other (vector)
central charges. The unique positive energy operator is the timelike
component of the translations in the ordinary type IIA and IIB theories,
but in other versions (*-theories or theories with a different
signature), it is not the $P_0$ operator that is positive, but rather a
component of a central charge operator.
\medskip
\section*{Acknowledgments.}
We have enjoyed a very stimulating atmosphere at the Memorial conference
for prof. G\"{u}rsey in Istanbul,
and thank the organizers for providing these opportunities.
This work was supported by the European Commission TMR programme
ERBFMRX-CT96-0045, in which E.B. is associated with Utrecht University.

\end{document}